\documentclass[twocolumn]{article}
\usepackage[dvips]{graphicx}
\usepackage{amssymb}

\newcommand{\eps} {\varepsilon}

\begin{document}

\title{Electrostatic Cancellation of Gravity Effects in Liquid 
Mixtures}\author{Yoav Tsori$^1$ and Ludwik Leibler$^2$\\~\\
$^1$ Physique de la Mati\`{e}re Condens\'{e}e, Coll\`{e}ge de France,
Paris, France\\
$^2$ Laboratoire Mati\`ere Molle \& Chimie (UMR 167)\\
ESPCI, 10 rue Vauquelin, 75231 Paris CEDEX 05, France}

\date{21/1/2005}

\maketitle
PACS: 05.20.-y, 64.70.Ja

\begin{abstract}
\footnotesize{
We point out that a spatially-varying electric field can be used
to cancel the effect of gravity in liquid mixtures by coupling to
the different components' permittivities. Cancellation occurs if
the system under consideration is small enough. For a simple
``wedge'' electrode geometry we show that the required system size
and voltage are practical, easily realizable in the Lab. Thus this
setup might be a simple alternative to more expensive or
hazardous options such as the space-shuttle, drop-tower, or magnetic 
levitation experiments.}\\~\\
\end{abstract}

The gravitational force brings about unwanted effects in many
experiments where phase-transitions are studied. Buoyancy effects
in liquid mixtures and colloidal suspensions, for example, become
increasingly important close to the critical point. 

A common
approach to negate gravity is to use a rocket, a plane or a
drop-tower facility \cite{space}. The space-shuttle is no longer
an alternative as NASA's new mission is to go to Mars rather than
do microgravity experiments. This approach suffers from many
shortcomings such as price, risk to human life, etc. A second
approach consists of using magnetic levitation  to compensate
gravity forces \cite{beysens}. In this case too the experimental
setup, based on a superconductor, is complicated and expensive.
\begin{figure}[h!]
\begin{center}
\includegraphics[scale=0.3,bb=125 165 425 705,clip]{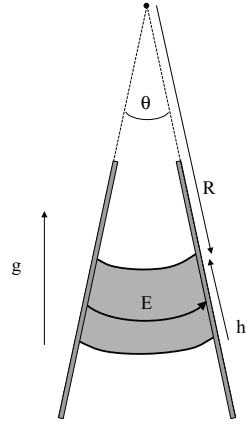}
\end{center}
\caption{{\small Illustration of the suggested geometry. Two flat
electrodes
with voltage difference $V$ are
tilted with an angle $\theta$. In the mixture (shaded
region), situated at distance $R$ from the imaginary meeting point, 
thefield
$E=V/\theta r$ is azimuthal. $h\ll R$ is the height above a fixed
reference point.
The setup shown assumes $\rho_A>\rho_B$ and $\eps_A>\eps_B$,
otherwise it should be turned up side down.}
}
\end{figure}

We propose here an alternative based on negating gravity forces by 
using electrostatics forces in a wedge geometry.
We consider for simplicity a binary mixture of two liquids A and B
with dielectric
constants $\eps_A$ and $\eps_B$ and densities $\rho_A$ and $\rho_B$,
respectively. The critical exponents, after the gravity effect had 
been negated,
are the same for all systems in the same (Ising) universality class.
The coupling of the gravitational force to the densities of the two
liquids
contributes a free-energy density
\begin{eqnarray}\label{Fg}
F_g=\left(\rho_0+(\rho_A-\rho_B)\phi\right)gh
\end{eqnarray}
Here, $g$ is the gravitational acceleration, $\phi$ is the local
A-component mixture composition ($0<\phi<1$), $h$ is a height
above some fixed reference and the mixture density $\rho(\phi)$ is
given by a linear relation,
$\rho(\phi)=\rho_0+(\rho_A-\rho_B)\phi$.

As pointed out before \cite{voronel,moldover}, a field varying
like $E\sim\sqrt{h}$ is needed to cancel gravity, because $E^2\sim
h$ is linear, just like the gravitational field. On the other hand,
$E^2$ always varies linearly if the spatial extent under consideration 
is small enough, and here we suggest one geometry where the linear
dependence on the spatial coordinate can be used in a practical
device.

The liquid mixture should be confined in an apparatus consisting
of two ``wedge'' shaped electrodes, see Figure 1. The electric
field then couples to the different dielectric constant of the
components and can counteract gravity. In a uniform medium, the
electric field {\bf E} points in the azimuthal direction and its
amplitude is
\begin{eqnarray}
E=V/\theta r
\end{eqnarray}
where $\theta$ is the opening angle of the wedge, $r$ is the
distance from the origin and $V$ the potential difference between
the two plates. The electrostatic contribution to the free-energy
density of the mixture is given by $F_{\rm es}=-\frac{1}{8\pi}\eps{\bf
E}^2$. For small density variations, as is relevant close to a
critical point, one can expand $\eps(\phi)$ to linear order in
$\phi$: $\eps=\eps_0+(\eps_A-\eps_B)\phi$.

In equilibrium, density variations will be azimuthal (they will
depend on $r$ only), since any deviation from such a density
profile costs energy \cite{AH,TA}. In this case, the solution of
Laplace's equation for the field is still $E=V/\theta r$. In the
experiment, the system should be confined to a small region whose
extension is much smaller than $R$ and the angle $\theta$ should
be small. In this case it is possible to expand $r$ around $R$ and
obtain the electrostatic contribution to the free-energy
\begin{eqnarray}\label{Fes}
F_{\rm es}\simeq -\frac{1}{4\pi} \left(\eps_0+(\eps_A-\eps_B)\phi\right)
\frac{V^2}{\theta^2 R^3}h+const.
\end{eqnarray}
Hence, considering terms linear in $\phi$ in Eqs. (\ref{Fg}) and
(\ref{Fes}),
one finds that the electrostatic and gravitational free-energy 
densities exactly
cancel throughout the whole sample volume if
\begin{eqnarray}
\left(\rho_A-\rho_B\right)g=\frac{1}{4\pi}\left(\eps_A-\eps_B\right)
\frac{V^2}
{\theta^2R^3}
\end{eqnarray}

Let us examine the numerical values of the parameters required for
this method to work. For concreteness we consider a 
methanol/cyclohexane mixture, where the density difference is
$\rho_A-\rho_B\simeq 130$ Kg/m$^3$, and the permittivity difference is 
$\eps_A-\eps_B\simeq 31.6\times 8.9\cdot 10^{-12}$ F/m. At a voltage 
of $V=100$ V and opening angle $\theta=5$ degrees, we find that 
$R=0.6$ cm. This value of $R$ means that the system itself (with 
dimensions $h\ll R$) can not be larger than few millimeters. However, 
decreasing the opening angle to $1$ degree and increasing $V$ to $500$ 
V shows that $R$ is much larger, $R=5.2$ cm, allowing for a 
correspondingly larger system size. The corresponding fields for the 
two cases above are $E=V/\theta R=1.9\cdot 10^5$ V/m and $E=5.5\cdot 
10^5$ V/m, well below dielectric breakdown. At these rather low 
fields, the field-induced shift of the critical 
temperature is found to be 
negligible (smaller than $3\cdot 10^{-4}$ K)
\cite{TTLnature,orz}.

The device proposed by us can thus be easily realized in the lab.
 In order to avoid problems of charge injection it would be
advised to use moderate frequency ($\gtrsim 1$ kHz) AC fields. The 
device suggested (and other geometric variants)
has many advantages over air-borne and magnetic devices, the most
obvious ones are the simplicity and small price of the setup.

\end{document}